# State of Security Awareness in the AM Industry: 2020 Survey

Mark Yampolskiy[1], Paul Bates[2], Mohsen Seifi[3], and Nima Shamsaei[4]

**ABSTRACT**

Security of Additive Manufacturing (AM) gets increased attention due to the growing proliferation and adoption of AM in a variety of applications and business models. However, there is a significant disconnect between AM community focused on manufacturing and AM Security community focused on securing this highly computerized manufacturing technology. To bridge this gap, we surveyed the America Makes AM community, asking in total eleven AM security-related questions aiming to discover the existing concerns, posture, and expectations.

The first set of questions aimed to discover how many of these organizations use AM, outsource AM, or provide AM as a service. Then we asked about biggest security concerns as well as about assessment of who the potential adversaries might be and their motivation for attack. We then proceeded with questions on any experienced security incidents, if any security risk assessment was conducted, and if the participants' organizations were partnering with external experts to secure AM. Lastly, we asked whether security measures are implemented at all and, if yes, whether they fall under the general cyber-security category.

---

[1] Department of Computer Science and Software Engineering, Auburn Cyber Research Center (ACRC), and National Center for Additive Manufacturing Excellence (NCAME), Auburn University, Auburn, AL, 36849, U.S.; https://orcid.org/0000-0003-4626-2754
[2] ASTM International, Washington, DC, 20036, U.S., https://orcid.org/0000-0002-5591-6820
[3] ASTM International, Washington, DC, 20036, U.S.; https://orcid.org/0000-0001-8385-2337
[4] Department of Mechanical Engineering and National Center for Additive Manufacturing Excellence (NCAME), Auburn University, Auburn, AL, 36849, U.S.; https://orcid.org/0000-0003-0325-7314
Corresponding Author: Mark Yampolskiy, mark.yampolskiy@auburn.edu

Out of 69 participants affiliated with commercial industry, agencies, and academia, 53 have completed the entire survey. This paper presents the results of this survey, as well as provides our assessment of the AM Security posture. The answers are a mixture of what we could label as expected, "shocking but not surprising," and completely unexpected. Assuming that the provided answers are somewhat representative to the current state of the AM industry, we conclude that the industry is not ready to prevent or detect AM-specific attacks that have been demonstrated in the research literature.

Introduction

The broad adoption of Additive Manufacturing (AM) became a matter of fact and does not require convincing anymore. While not even ten years ago, at AM conferences one could frequently hear questions like "what could be the good use-case for AM besides the famous GE's fuel injection nozzle?" or arguing in favor like "AM gives us complexity for free," today the conversation shifted to two distinct but technical subjects. One, how to resolve the remaining technological challenges and allow AM to live to its fullest potential. Another, however, reflects growing concern that AM will become target of (what is often referred as) "cyber-security attacks." This paper is dedicated solely to the latter.

Despite increased awareness of security, AM is not unique in the way how security risks are ignored by the technology developers and adopters. Just to name a few examples, we have seen security being neglected in industrial infrastructure (or SCADA systems) until the famous

Stuxnet attack[1] demonstrated its vulnerability, and in modern cars until they were famously hacked and remotely taken control of.[2,3] In these examples, the security might have been explained away by the fact that previously secure systems became increasingly computerized and networked. However, when the natively computerized and networked Internet of Things (IoT) devices were introduced, they went through the same phases of technology development, broad adoption, discovery and demonstration of security vulnerabilities, and only then development and implementation of the security measures.

As of mid-2020, AM industry is clearly in the phase of a broad technology adoption. While there have not been any broadly publicized reports about real AM-specific attacks, security incidents like ransomware that does not spare even hospitals became almost a weekly news. Thus, no wonder that AM industry becomes increasingly concerned about security.

While the concern about potential of upcoming attacks is growing in the AM industry, it is not clear to what extent it is ready to face real attacks. To assess this, we conducted a survey; its results are presented in this paper. In the remainder of the paper, we first present the survey questions and characterize the participants. We then present answers to individual questions alongside the related preliminary analysis rooted in AM Security literature. We conclude this paper with our overall assessment of the state of awareness and preparedness.

Survey Questions and Participants

The AM Security field was pioneered around 2014,[4] with both attacks and defense measures discussed in the research literature since then. We prepared a set of questions based on the understanding of the problem space developed in the field. At the same time, we formulated the questions trying to avoid implicit bias and to provide respondents with the ability to provide answers that not necessarily would fit in our understanding of the problem space. Some questions consisted of multiple parts; only if part "a" was answered in a certain way (mostly positively), the follow-up question(s) were asked. The prepared questions are listed in Table 1.

**TABLE 1 - QUESTIONS USED IN THE SURVEY**

| Q# | Question |
|---|---|
| Q1 | Does your organization manufacture with AM? |
| Q2 | What related security risks/concerns have you identified related to your manufacturing using AM? |
| Q3.a | Does your organization outsource AM? |
| Q3.b | What related security risks/concerns have you identified related to outsourcing your AM? |
| Q4.a | Does your organization provide AM manufacturing services to external companies/agencies/individuals? |
| Q4.b | What related security risks/concerns have you identified as it relates to providing your AM manufacturing services? |
| Q5 | What are your biggest security concerns associated with additive manufacturing (AM)? |
| Q6 | Who are potential attackers/adversaries you expect would attack AM in your company/agency? |
| Q7 | What do you assess would be their goals/objectives? |
| Q8.a | Has your organization experienced a cyber incident related to AM activities? |
| Q8.b | Please outline your cyber incident experience (if possible). |
| Q9.a | Has your organization partnered with external AM security providers? |
| Q9.b | Which security concerns do you have with using these external AM security providers? |
| Q10.a | Has your organization conducted a formal or informal risk assessment related to AM? |
| Q10.b | Which risks, related to AM, have been identified during this assessment? |
| **Q11.a** | Does your organization have a security program in place for AM? |
| **Q11.b** | Does it fall under the general cyber-security? |
| **Q11.c** | How does it go beyond general cyber security? |

We distributed a call for participation in the survey through several mailing lists. Overall, 69 participants answered the call. They are affiliated with 64 unique organizations, representing industry, governmental agencies, research laboratories, and universities (see Appendix A for a curated list of participants' affiliations). As it is common with the surveys, only 53 out of 69 participants answered all questions, what correspond to 77%.

Survey Results and Per-Question Analysis

In this section we provide answers to individual questions. For each of the results, we provide a preliminary analysis rooted in the AM Security literature. Our conclusions only apply to organizations whose members participated in the survey. We advise against deriving more wider-reaching conclusions, because (i) the number of participants is relatively small compared to the number of organizations overall, (ii) the knowledge and understanding shown by the survey participants does not necessarily represents the institutional knowledge of their organizations, and (iii) as the majority of participants are affiliated with the U.S. organizations, breakdowns for other countries can look completely different.

**Q1: DOES YOUR ORGANIZATION MANUFACTURE WITH AM?**

All 69 (100%) respondents answered the question. 52 (75%) of them answered "yes" and 17: (25%) "no" (see **fig. 1**).

Page 5 of 35

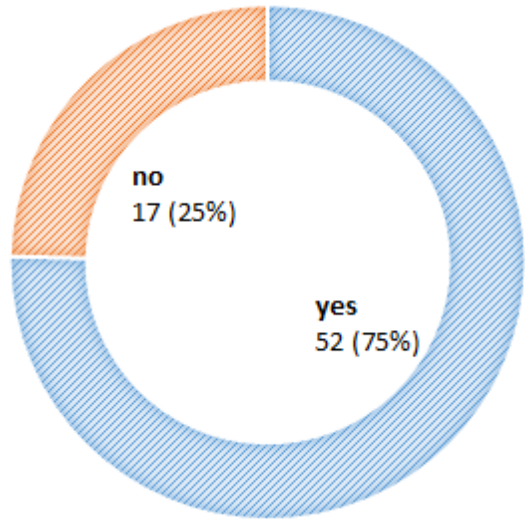

FIGURE 1 - ANSWERS TO Q1

This indicates that among organizations with interest to AM, a high percentage already uses this technology.

## Q2: WHAT RELATED SECURITY RISKS/CONCERNS HAVE YOU IDENTIFIED RELATED TO YOUR MANUFACTURING USING AM?

Participants were given a free-form field to answer. Only 37 participants (corresponds to 53%) answered this question. By analyzing the answers, we have seen that while some of them could be expected, there were also some unexpected and even disturbing. Below we provide some examples.

Provided Answers – Expected (excerpt):

- *"Uncontrolled access to intellectual property (designs)"*
- *"Leakage of manufacturing data"*
- *"Cloud based data capture for access from remote R&D facility overseas"*
- *"Part build files: Intentional and unintentional (Tampering, translational errors*

*between tools)"*

- *"Corruption of AM design files by unexpected network activity"*
- *"Data corruption and lack of security on data transfer from storage systems to printers"*

Provided Answers – Unexpected, sometimes Shocking (excerpt):

- *"None"*
- *"None that I am aware of"*
- *"We buy IT services that are supposedly at NIST 800 -171 and will meet CMMC level 3 or more."*

For both these groups conclusions could be drawn. While the expected answers identify security threats discussed in the AM Security literature, the way how these are described reveals that the respondents are not familiar with the terminology used in the field; this might indicate lack of familiarity with the literature. While several peer-reviewed publications surveyed the AM Security research literature[4-10] as well as several partly overlapping taxonomies to structure the field have been proposed,[4,10,11] one could argue that we should not expect people without security backgrounds to read the AM security research literature. To bridge this gap, in authors' opinion, we need standards that are aligned with research results and provide AM industry clear recommendations and best practices how to secure AM.

The unexpected answers show presence of ignorance to the potential security issues and in overconfidence on a certification that was developed without specifics of AM Security in mind. In a particular instance, we want to note that the CMMC rulemaking process is currently ongoing; therefore, it is meaningless at this point to say that anything will "meet CMMC level 3 or more." The research literature has shown that even security solutions developed for subtractive manufacturing with CNC machines in mind would not necessarily work for AM,[12-13] and that established cyber-security considerations such as CIA Triad (here, CIA stands for confidentiality, integrity, availability) are applicable to AM only in certain cases while not applicable at all in many others.[14]

## Q3.A: DOES YOUR ORGANIZATION OUTSOURCE AM?

Only 61 (88%) respondents answered the question. 32 (52%) of them answered "yes" and 29 (48%) "no" (see fig. 2).

**FIGURE 2 - ANSWERS TO Q3.A**

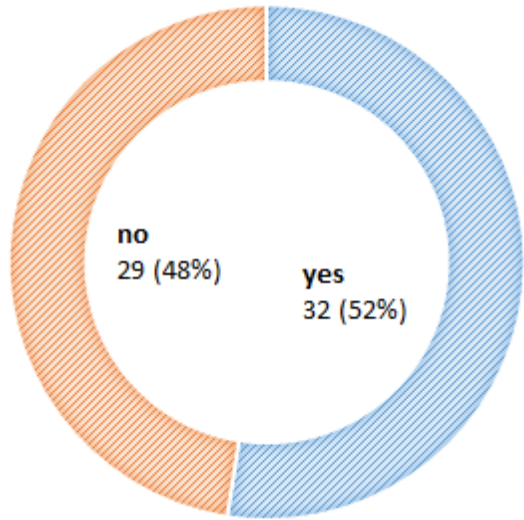

This shows that a slim majority of organizations consider that benefits of outsourcing outweigh whatever concerns they might have in conjunction with this business model.

## Q3.B: WHAT RELATED SECURITY RISKS/CONCERNS HAVE YOU IDENTIFIED RELATED TO OUTSOURCING YOUR AM?

Only 23 (33%) of participants answered this question. Also here participants were given an opportunity to provide a free-form answer. As with the prior question, we could distinguish between rather expected and unexpected answers. Below we list examples for both.

Provided Answers – Expected (excerpt):

- *"Data and IP loss / IP and loss of critical information"* (here, IP stands for Intellectual Property)
- *"Protection for CAD files, and confidential design and manufacturing data with email communication"*
- *"Used on the correct machine the correct number of times, especially if there is a failure or restart needed"*
- *"Sending files - used on the correct machine the correct number of times, especially if there is a failure or restart needed. ensuring no changes have been made once files are out of our hands"*
- *"How outsource vendors handle the information received"*
- *"In-situ monitoring can allow access to process parameters and scan strategies"*

Provided Answers – Unexpected, sometimes Shocking (excerpt):

- *"None. Any work outsourced has proper security safeguards"*
- *"Lack of standards in managing and transfer of proprietary CAD and build files"*
- *"Government requirements for data storage and transmission"*
- *"Sharing technical data files via email and not using a secure file share server"*

The expected answers show at least some degree of the problem space understanding. Most of the answers in this category indicate concerns about technical data theft. However, the way how answers are formulated indicate a mix-up between security threats (i.e., adversarial goal that can be achieved) and attack means (i.e., methods that an adversary could use to achieve these goals). Furthermore, like our analysis of the answers to Q2, most of the expected answers were given without using the terminology established in the research literature; this also indicates the lack of familiarity with the latter.

The unexpected answers reveal several deeper problems. The first answer in this category indicates possible overconfidence on undisclosed "security safeguards". The next two answers are of technical and not security nature; this indicates a fundamental misunderstanding of what security is responsible for. The last answer (sharing data files via e-mails) indicates that even established and easily available cyber-security mechanisms are not necessarily deployed and used. This is of particular concern, because it shows that an attacker can be successful even by using broadly available *exploits* (i.e., compromise methods in cyber domain).

## Q4.A: DOES YOUR ORGANIZATION PROVIDE AM MANUFACTURING SERVICES TO EXTERNAL COMPANIES/AGENCIES/INDIVIDUALS?

This question was a natural counterpart to the question Q3.a. Like with Q3.a, 61 (88%) of respondents answered the question. Here, however, 36 (59%) of them answered "yes" and 25 (41%) "no" (see fig. 3).

**FIGURE 3 - ANSWERS TO Q4.A**

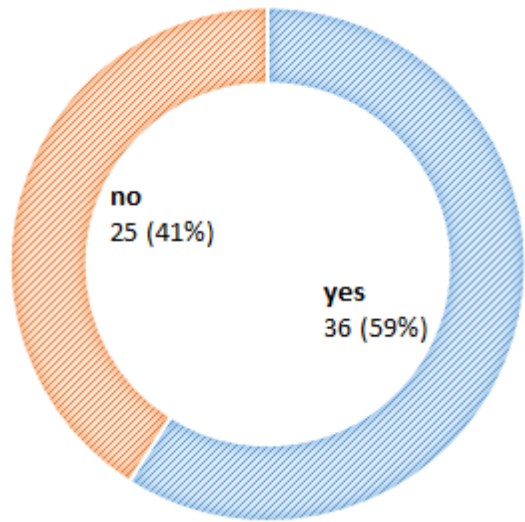

Slightly more organizations find it beneficial to provide AM as a service than using it as a service. Furthermore, 16 (26%) respondents answered positively to both Q3.a and Q4.a. This is an interesting and rather unexpected result, because it shows that a substantial portion of organizations might be involved in AM outsourcing model on both sides.

## Q4.B: WHAT RELATED SECURITY RISKS/CONCERNS HAVE YOU IDENTIFIED AS IT RELATES TO PROVIDING YOUR AM MANUFACTURING SERVICES?

Only 27 (39%) respondents whose organization provides AM as a service answered this question. We again distinguish between expected and unexpected answers.

Provided Answers – Expected (excerpt):

- *"Process theft as well as specific design theft / Design data copying"*
- *"For certification purposes some customer might require lot of details concerning the production process"*
- *"Counterfeit"*
- *"Export control concerns. In the healthcare space, protected health information (PHI)"*
- *"We transfer files over emails sometimes which isn't the safest/most efficient"*
- *"Potential malware handling electronic files"*

Provided Answers – Unexpected, sometimes Shocking (excerpt):

- *"None / n/a"*
- *"Our IT is outsourced and supposedly meets NIST 800-171"*
- *"Issues w/ unprepared supply chains related to traceability"*
- *"Lack of compatibility and difficulty in integration of data systems between*

*clients and internal systems"*

Like with the answers to Q3.b, the expected answers indicate unfamiliarity with the AM Security research literature and the terminology established in the field. The first two answers (talking about process parameters) indicate understanding that technical data in the AM context is not limited to the design files alone. The second answer shows a confusion between this security threat (i.e., the undesirable outcome of an attack) and the attack means (in this case, non-technical: customer "requiring lot of details"). The third (counterfeit) and fourth (export control) indicate valid concerns, which, however, are of legal and not of technical nature. The last two selected answers in this category indicate the negligence of using established cyber-security practices and tools; as in assessment to Q3.b, this enables use of broadly available exploits and hacker tools.

Among the unexpected answers, the first two indicate either complete overconfidence of security or overreliance of the certification, oblivious of fact that the mentioned certification was developed to address different aspects that are security concern in the AM space. In the mentioned example it is even worse than that: NIST SP 800-171 is a guidance document, not a regulation. Currently, anyone can claim that they comply with it. No formal certification authority exists (yet). The recently released CMMC 2,0 allows most DoD contractors to self-certify. The third unexpected answer shows a mix-up between AM and Supply Chain Securities; both are necessary but very distinct disciplines. It is indeed a controversial and not a clear-cut distinction when AM is being outsourced: does that makes the supply chain relationship an AM security concern? In any case, securing a business process is no less important than securing a manufacturing process. The last example answer listed as unexpected demonstrates misunderstanding of security concerns at a more fundamental level, because it talks about an issue that is not of security nature.

## Q5: WHAT ARE YOUR BIGGEST SECURITY CONCERNS ASSOCIATED WITH ADDITIVE MANUFACTURING (AM)?

After asking in Q3 and Q4 questions regarding security concerns in specific business models, with this question we took a broader perspective. The respondents could both select one of the pre-defined answers as well as add an additional answer in a free form field. Overall, 56 (81%) survey participants answered this question.

Selected Predefined Answer(s):

- *"Technical data theft (e.g., Design data, Process parameters, IP loss, Sensor information)"*: 50 (89% of all who answered the question)
- *"Sabotage (of manufactured part, AM equipment, environment)"*: 23 (41% of all who answered the question)

Other (freeform):

- *"Data integrity"*
- *"Maintaining data integrity through the entire process will be a challenge. Unintentional data corruption is likely a higher risk than hacking or theft."*
- *"Liability for tech data theft"*
- *"Legal risk such as if data is ITAR"*
- *"Violation of export control restrictions, violation of rules regarding critical infrastructures."*
- *"Traceability of material sources"*

It was no surprise that the vast majority of respondents who answered the question (89%)

selected Technical Data Theft as one of their security concerns. In the case of companies, this could lead to monetary impact on the business; in the case of governmental agencies and sometimes labs, such an attack could compromise national security. What was interesting is that also a substantial portion of respondents (41% of all who answered the question, to be exact) also consider intentional Sabotage as a real concern. While sabotage has been intensively studied in the research literature,[15-24] we were surprised to see that it is also considered a real concern by a substantial portion of surveyed organizations.

The freeform answers were both revealing and again concerning. The first two answers mixes up potential attack means (violation of data integrity) with a security threat (in this case, sabotage) to which it potentially could lead. The next three answers indicate rather legal than technical security concerns. The last answer indicates once again misunderstanding of what boundaries of security are: security disciplines are generally concerned about intentional actions taken by malicious actors, but not about solving technical issues under benign operational conditions.

## Q6: WHO ARE POTENTIAL ATTACKERS/ADVERSARIES YOU EXPECT WOULD ATTACK AM IN YOUR COMPANY/AGENCY?

56 (81%) of all respondents answered this question. As with the previous question, we have provided both predefined answers (multiple could be selected) as well as a field for the freeform answer.

Selected Predefined Answer(s):
- *"State actors"*: 29 (52%)
- *"Competitors"*: 39 (70%)

- *"Insider threat"*: 22 (39%)

Other (freeform):

- *"Hackers"*

- *"Script kiddies, environmental activists"*

- *"Individual threat actors looking to steal IP or hold our company hostage for ransom"*

- *"Ransomware"*

- *"Vendors and suppliers - IP loss"*

- *"Subcontractors can (and do) share our designs with competitors who use the same subcontractors"*

- *"Customers. Some without ethics will also steal IP"*

- *"Could be issues with our customers' competitors"*

- *"Actors wanting to compromise the integrity of transportation infrastructure"*

From the predefined answers, we were surprised to see that most respondents answered this question (70%) selected the "competitors" option. As AM Security literature is dominated by publications from academia, we think that the breakdown between the predefined options can provide a more realistic input for the future field development.

From the freeform answers, we found especially interesting mentioning about "vendors", "subcontractors", and "customers". This indicates a concern (regardless of whether it is realistic or just perceived) that entities involved in the contractual relationships can be malicious. This could be another valuable input to the AM Security field: While malicious actors have been considered,[25-26] most of the research literature is dominated by the considerations of attacks conducted by an external adversary.

## Q7: WHAT DO YOU ASSESS WOULD BE THEIR GOALS/OBJECTIVES?

56 (81% of all respondents) answered this follow-up question.

Selected Predefined Answer(s):

- *"Exploratory/reconnaissance (e.g., intelligence gathering)"*: 47 (84%)
- *"Monetary gains (e.g., through ransomware)"*: 23 (41%)
- *"Patent infringement"*: 33 (59%)
- *"Denial of service (e.g., delay product launch)"*: 16 (29%)

Other (freeform):

- *"Negligence (no particular objective)"*
- *"Threat from student using equipment"*
- *"Stealing customer's IP and drawings"*
- *"Steal trade secrets"*
- *"Embed defects in parts / Introduce critical flaws into parts to cause catastrophic failures"*
- *"Destruction of equipment, i.e., printing faulty parts that aren't detectable"*
- *"Damage to transportation infrastructure […] inserting undetectable flaws in parts ordered so they might"*

Interestingly, the majority of respondents answered this question (84%) selected the predefined option "exploratory/reconnaissance". Several non-invasive *side-channel* attacks have been shown in the research literature[26-29] that would definitely suffice this purpose but not necessarily would provide quality sufficient for part infringement.

While most of the freeform answers are of technical and security-related nature, the first two are definitely not. Once again, it indicates that the responsibility of security is not necessarily clear to the survey participants, and possible to their organizations. At the same time, we need to point out that some security controls, such as separation of duties and least functionality, have ancillary non-security benefits. Thus, robust security should not be seen by AM industry just as an expensive necessity or as a mandate, but also as means of protecting against costly mistakes, human error, etc.

## Q8.A: HAS YOUR ORGANIZATION EXPERIENCED A CYBER INCIDENT RELATED TO AM ACTIVITIES?

As no real AM-specific attacks have been publicized, this question was intended to explore how realistic are concerns discussed in AM Security community. 55 (80%) respondents answered this question. 5 of them (9%) answered "yes" and 50 (91%) answered "no" (see fig. 4).

**FIGURE 4 - ANSWERS TO Q8.A**

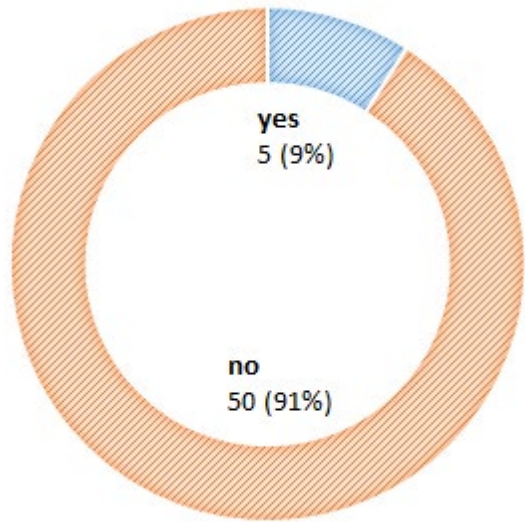

This demonstrate that real AM-specific attacks have already been conducted. Furthermore, the actual number could have been higher than 9%, because of several reasons. For example, participants might not be aware of all attacks their organization has been exposed; even if they are, they might have decided not to disclose this potentially sensitive information.

### Q8.B: PLEASE OUTLINE YOUR CYBER INCIDENT EXPERIENCE (IF POSSIBLE).

Only respondents who answered positively Q8.a have been asked this question. The answer was a freeform field. Only two (2) of the participants answered this question, but after looking at their replies we had to correct participation to 0%.

Answers given (freeform):

- *"no comment."*
- *"n/a"*

This indicates lack of willingness to share information. This is even more concerning in combination with the answer to Q8.a, because it indicates the posture in AM industry to face security challenges alone (without collaboration and information sharing with other similar entities). Other non-native security sectors (e.g., critical infrastructure) had went through a similar development till broader realization that such posture does not serve security of anybody. Meanwhile, there are serious attempts to mandate incident reporting,[30] thus enabling means of collective protection.

### Q9.A: HAS YOUR ORGANIZATION PARTNERED WITH EXTERNAL AM SECURITY PROVIDERS?

This was the first in a series of questions aiming to identify how security concerns are addressed. 54 (78%) of participants answered this question, 12 (22% of those who answered) selected "yes" and 42 (78%) "no" (see fig. 5).

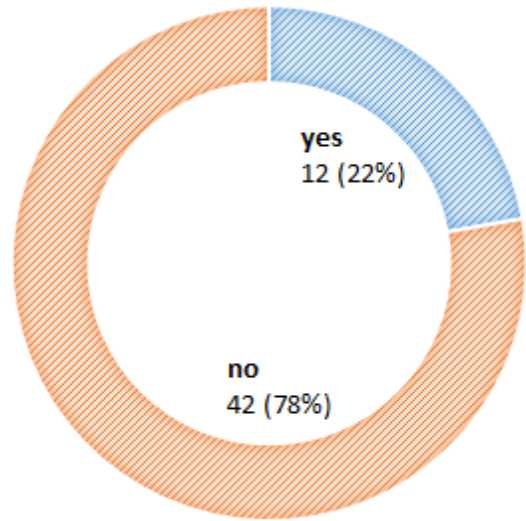

FIGURE 5 - ANSWERS TO Q9.A

yes 12 (22%)
no 42 (78%)

The answers indicate that most surveyed organizations do not partner with AM Security professionals. The reasons for this reluctance present an interesting question; unfortunately, it was not explored in this survey. Nevertheless, this also indicates a business opportunity for AM-specialized security products.

## Q9.B: WHICH SECURITY CONCERNS DO YOU HAVE WITH USING THESE EXTERNAL AM SECURITY PROVIDERS?

This question was presented only to respondents who answered Q8.a positively. We received 6 (or 50%) answers.

Answers (freeform):

- *"Don't rely on external security expertise but build up your own expertise"*
- *"All external activity is a risk. It means sharing more information with outsiders which makes it easier for an attacker to enter through another channel"*
- *"While they are there to help, they also create another potential door into our data systems, for a team that we don't know and we don't have the ability to vet,* *beyond doing what due diligence we can in deciding to work with the company. [...] we are AM & materials experts, not IT experts, so this raises question - how do we monitor the external IT security team? When looking at recent events, like the Treasury Dept breach, it's quite a daunting question for any small companies when this got past large security providers and entire federal agencies!"*

- *"Security approach is not wholistic; tends to be focused on protecting design IP"*
- *"We trust them but I'm not sure how we'd audit them since IT is not our area of expertise. The right things seem to be in place like dual VPN"*
- *"We are engaging a CMMC consultant so not sure if that applies"*

The answers show that even in the cases when an external security company is involved, the AM company might have concerns about its security or trustworthiness. This might also indicate at least some of the reasons why AM organizations are reluctant to partner with external security providers (as per answers to Q9.a). The cause of this can be a lack of standards that could provide means of assurance that a security products and services provided by third parties are trustworthy.

## Q10.A: HAS YOUR ORGANIZATION CONDUCTED A FORMAL OR INFORMAL RISK ASSESSMENT RELATED TO AM?

53 (77%) or all participants answered this question, 20 (38%) of them answered "yes" and 33 (62%) "no" (see fig. 6).

**FIGURE 6 - ANSWERS TO Q10.A**

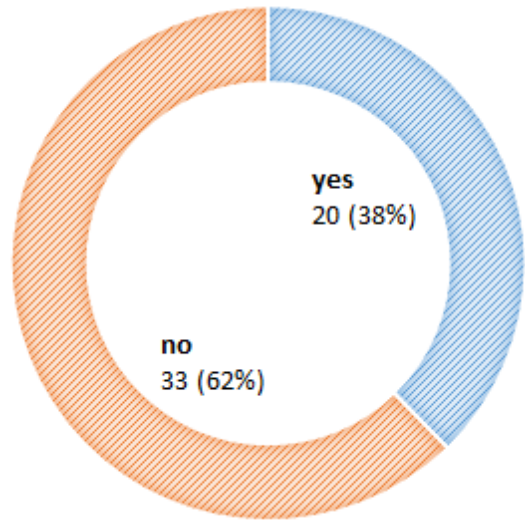

This result is very disconcerting, because it shows that the majority of surveyed organizations (62%) don't appear to take security issues seriously. This indicates a wide-spread reactive mindset of "if it is not mandated, I don't worry about it."

## Q10.B: WHICH RISKS, RELATED TO AM, HAVE BEEN IDENTIFIED DURING THIS ASSESSMENT?

This was the follow-up that was only asked when Q10.a was answered positively. We received 13 answers, but after discarding answers like "Prefer not to say" and "no comment" we only left with 9 (45% of those who were asked this question).

Answers (freeform):

- *"Data and IP loss through insider access to programming. Outside access restricted by company IT policies."*
- *"IP loss in engaging AM service bureaus and suppliers Data security in transfer of files within and to vendors"*
- *"Potential 'reverse engineering' of processing data to gain intelligence of*

- *products manufactured"*
- *"Inadvertent sharing of sensitive data (we are a global company)"*
- *"Loss of IP due to data theft, 2. Damage to printers that are connected to the internet"*
- *"Intentional corruption of CAD Software, build files, tool files. Loss of intellectual property"*
- *"Misconfigurations, account hacking, insider threat, weak control plane, insecure interfaces"*
- *"Data encryption"*
- *"Data loss, impact on company reputation, production of defective parts, revenue loss"*
- *"Data management, facility access, internal (employee) and external (IT service, IT security) access. We have gone through a full NIST Cybersecurity framework assessment and about to start a DoD NIST 800.171 Cybersecurity assessment."*

As with the prior questions, first six answers indicate general understanding of the security threats but use of "home-grown" explanations instead of the terminology established in the AM Security field; this indicates lack of knowledge of the literature. The seventh's answer ("data encryption") indicates a misunderstanding of what are security risks, security threats, and security techniques that can be used to counter those. The last answer, once again, indicates overconfidence rooted in a certification.

**Q11.A: DOES YOUR ORGANIZATION HAVE A SECURITY PROGRAM IN PLACE FOR AM?**

With this question, we tried to understand whether organizations try to address or ignore security threats associated with AM. 51 (74%) answered this question, 21 (41%) of them "yes" and 30 (59%) "no" (see fig. 7)

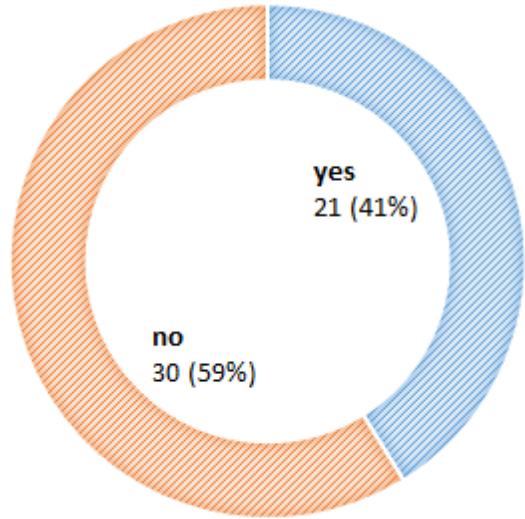

FIGURE 7 - ANSWERS TO Q11.A

The answer indicates that the majority (59%) of the surveyed organizations simple ignore security of AM.

### Q11.B: DOES IT FALL UNDER THE GENERAL CYBER-SECURITY?

Only respondents who answered Q11.a positively were asked this question. All 21 (100% of them) answered this question, 19 (90%) with "yes" and 2 (10%) with "no" (see fig. 8).

**FIGURE 8 - ANSWERS TO Q11.B**

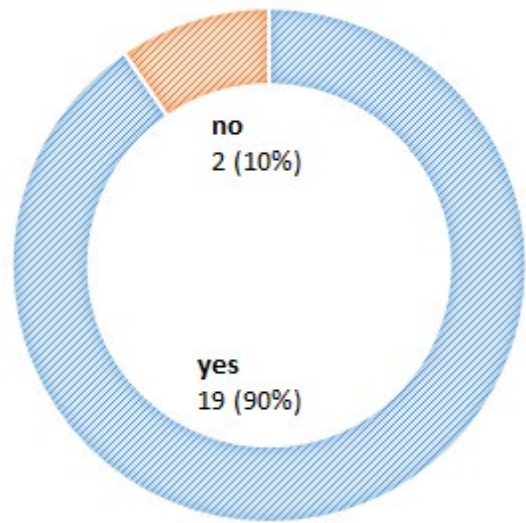

While not surprising, it is also concerning. In the AM Security literature, numerous publications demonstrated cyber-physical attack methods capable of bypassing cyber-security measures,[26-29] as well as cyber-physical and physical defense measures were proposed.[31-46] Also here, a need for a specific standard becomes obvious: derived from a constantly growing list of scientific publications, a curated view of possible attack methods and proposed countermeasures alongside with their respective limitations should provide companies with means for a more realistic assessment of their AM security.

**Q11.C: HOW DOES IT GO BEYOND GENERAL CYBER SECURITY?**

Only the respondents who answered Q11.b with "no" were asked also this question. None of them (0%) provided an answer.

This shows unwillingness to share information that might have contributed to the improved security posture in AM industry.

Summarizing our preliminary analysis of answers to individual questions, we conclude that the results are "shocking but not surprising." Often, we have observed the lack of the problem space understanding. It took several forms, such as frequent mix-ups of security threats, technical means that attacks can use, and impact on the business. Furthermore, some organizations substitute AM Security through compliance with frameworks developed with different use cases, technologies, and threat models in mind. As mentioned throughout the text, we think that a dedicated AM Security standard or guidelines (none of which exists now) could significantly improve this situation.

We have also seen the unwillingness to share information about AM-specific attacks, ignoring potential benefits of collective protection and causing instead the situation when every organization has to face the security threats alone. But even then, only 41% of the surveyed organizations have a dedicated AM Security program in place, but 90% of those treats cyber-security as a sole solution to all problems – a posture dangerous in the view of various cyber-physical attacks already demonstrated in the AM Security literature.

Thus, we are coming to a very concerning conclusion that most organizations are NOT ready to adequately address AM Security.

**ACKNOWLEDGEMENTS**

This work was funded in part by the America Makes under the award 5001.005, and partially supported by the U.S. Department of Commerce, National Institute of Standards and Technology under Grants NIST-70NANB19H170 and NIST-70NANB20H193.This work was funded in part by the America Makes under the award 5001.005, and partially supported by the U.S. Department of Commerce, National Institute of Standards and Technology under Grants NIST-70NANB19H170 and NIST-70NANB20H193.

# Appendix A – Survey Participant's Affiliation

| PARTICIPANT'S AFFILIATION | ORGANIZATION TYPE |
|---|---|
| **3D Metalforge** | Industry |
| **3D Systems** | Industry |
| **Additive Industries North America** | Industry |
| **Ansys Inc.** | Industry |
| **Applied Research Laboratory at Penn State** | Research Laboratory |
| **Advanced Remanufacturing and Technology Centre (ARTC)** | Industry |
| **American Society for Testing and Materials (ASTM)** | Standardization Organization |
| **Battelle Memorial Institute** | Industry |
| **ICON** | Industry |
| **Buckeye Furniture** | Industry |
| **Cincinnati** | Industry |
| **ConocoPhillips** | Industry |
| **Coventry University,** <br> **TWI Ltd** | Academia, <br> Research Laboratory |
| **Defense & Energy Systems, LLC** | Industry |
| **The DoD Manufacturing Technology (ManTech)** | Research Laboratory |
| **diondo** | Industry |
| **EOS North America** | Industry |
| **EWI** | Research Laboratory |
| **Fabrisonic, LLC** | Industry |
| **Farcco Tecnologia Industrial** | Industry |
| **FormAlloy Technologies, Inc.** | Industry |
| **The Fraunhofer Research Institution for Additive Manufacturing Technologies IAPT (Fraunhofer IAPT)** | Research Laboratory |
| **General Electric (GE)** | Industry |
| **GL** | Industry |
| **Guaranteed B.V.** | Industry |
| **Honda R&D Americas** | Research Laboratory |
| **Intermecanic** | Industry |
| **Ivaldi Group** | Industry |
| **Johnson & Johnson** | Industry |
| **Konica Minolta, Inc.** | Industry |
| **KPMG** | Industry |
| **Leonardo Helicopters** | Industry |
| **Lithoz America, LLC** | Industry |
| **Lawrence Livermore National Laboratory (LLNL)** | Research Laboratory |
| **Massachusetts Department of Transportation** | Governmental Agency |
| **MatterHackers** | Industry |
| **Mighty Buildings** | Industry |
| **The National Aeronautics and Space Administration (NASA)** | Governmental Agency |
| **National Center for Defense Manufacturing and Machining** | Research Laboratory |
| **(NCDMM)** | |
| **America Makes - National Additive Manufacturing Innovation Institute** | Governmental Agency |
| **nCode Federal, LLC** | Industry |
| **North Carolina State University (NCSU)** | Academia |
| **Northrop Grumman Corporation (NGC)** | Industry |
| **NGK Spark Plug Co., Ltd** | Industry |
| **Northwestern University** | Academia |
| **NuVasive** | Industry |
| **Oak Ridge National Laboratory (ORNL)** | Research Laboratory |
| **The Ohio State University's Center for Design and Manufacturing Excellence (CDME)** | Research Laboratory |
| **Perrygo consulting** | Industry |
| **Raytheon Technologies Corporation** | Industry |
| **Raytheon Technologies Research Center** | Research Laboratory |
| **Red Fox Tavern** | Industry |
| **Renault Sport Racing** | Research Laboratory |
| **SAE International** | Standardization Organization |
| **Siemens Energy** | Industry |
| **TCI Finance Industry Association** | Industry |
| **The Boeing Company** | Industry |
| **The University of Dayton Research Institute (UDRI),** | Academia, |
| **Air Force Research Laboratory (AFRL)** | Research Laboratory |
| **UK Atomic Energy Authority** | Research Laboratory |
| **University of Toledo** | Academia |
| **USACE Information Technology Lab** | Research Laboratory |
| **The University of Texas at El Paso (UTEP)** | Academia |
| **Virgin Media** | Industry |
| **Wichita State University (WSU),** **National Institute for Aviation Research (NIAR)** | Academia, Research Laboratory |